\documentclass[a4paper, 12pt]{article}

\usepackage{a4wide}
\usepackage{graphicx}
\usepackage[latin1]{inputenc}
\usepackage[english, swedish]{babel}
\usepackage{amssymb}

\begin{document}

\setlength{\parskip}{1ex}
\setlength{\parindent}{0pt}
\selectlanguage{english}

TSL/ISV-2004-0277\\
June 2004\\

\begin{center}
{\Large \bf Interpretation of electron-proton scattering at low $Q^2$}\\
\bigskip\bigskip

{{\bf J.~Alwall$^a\,$}\footnote{E-mail: johan.alwall@tsl.uu.se}, 
{\bf G.~Ingelman$^{a,b}\,$}\footnote{E-mail: gunnar.ingelman@tsl.uu.se}} \\
\medskip 
{$^a$ High Energy Physics, Uppsala University, Box 535, S-75121 Uppsala, Sweden}\\
{$^b$ Deutsches Elektronen-Synchrotron DESY, D-22603 Hamburg, Germany}\\ 
\end{center}

\begin{abstract}
Low-$Q^2$ photons do not resolve partons in the proton, which gives problems when applying the deep inelastic scattering formalism, such as an unphysical, negative gluon density extracted from data. Considering instead hadronic fluctuations of the photon, we show that the generalised vector meson dominance model (GVDM) gives a good description of the measured cross section at low $Q^2$, {\it i.e.}\ reproduces $F_2(x,Q^2)$, using only few parameters with essentially known values. Combining GVDM and parton density functions makes it possible to obtain a good description of $F_2$ data over the whole range of $x$ and $Q^2$. 
\end{abstract}

\section{Introduction}
Experimental measurements on electron-proton ($ep$, and also $\mu p$) scattering are usually interpreted in terms of the theoretical formalism for deep inelastic scattering (DIS). The differential cross section is then expressed in terms of proton structure functions given by the density functions for different partons, {\it i.e.}\ $q(x,Q^2)$ and $g(x,Q^2)$ for quarks and gluons carrying a fraction $x$ of the proton's energy-momentum when probed with the scale $Q^2$. The structure function $F_2$, which gives the dominant contribution to the cross section, is in leading order given by $F_2(x,Q^2) = \sum_q e_q^2 \left( xq(x,Q^2) + x\bar{q}(x,Q^2)\right)$ while the gluon density enters indirectly via the logarithmic $Q^2$ dependence of perturbative QCD. 

This formalism has also been applied to $F_2$ data at low photon virtuality $Q^2$, where the exchanged photon is not far from being on-shell. Parametrising such $F_2$ data in terms of quark and gluon density functions results in gluon distributions that tend to be negative for small $x$ at small $Q^2$ ({\it e.g.}\ $x\sim 10^{-4}$, $Q^2\sim 2\, \rm{GeV}^2$) \cite{negative-gluon,mrst}. The reason for this is that the DGLAP evolution, driven primarily by the gluon at small $x$, otherwise gives too large parton densities and thereby a poor fit to $F_2$ in the genuine DIS region at large $Q^2$. Although one may argue that the gluon density is not a directly observable quantity and hence might be negative, it certainly is in conflict with the interpretation of the probability for a gluon with momentum fraction $x$ in the proton. In particular, such a gluon distribution could be just an effective description for a more proper theoretical understanding. It need not have the same universality as proper parton density functions, thus giving incorrect results when applied to other interactions. For example, differences in the predicted Higgs production cross section (dominated by $gg\rightarrow H$) at the Tevatron and LHC arise depending on whether the gluon parametrisation is forced to be positive definite or allowed to be negative at small $x$ \cite{mrst}.

In this Letter, we argue that the root of the problem is the application of the formalism for DIS also in the low-$Q^2$ region, where the momentum transfer is not large enough that the parton structure of the proton is clearly resolved. The smallest distance that can be resolved is basically given by the momentum transfer of the exchanged photon through $d=0.2/\sqrt{Q^2}$, where $d$ is in Fermi if $Q^2$ is in GeV$^2$. This indicates that partons are resolved only for $Q^2\gtrsim 1\, \rm{GeV}^2$. For $Q^2\lesssim 1\, \rm{GeV}^2$, there is no hard scale involved and a parton basis for the description is not justified. Instead, the interaction is here of a soft kind between the nearly on-shell photon and the proton. The cross section is then dominated by the process where the photon fluctuates into a virtual vector meson state which then interacts with the proton in a strong interaction. This is the essence of the vector meson dominance model (VDM), for a review see \cite{VDM}.

In the following we use the original generalised vector meson
dominance model (GVDM) \cite{GVDM} for $ep$ scattering at low
$Q^2$. We show that it gives a good description of the recent HERA
data extending the $Q^2$ region to very low values, which are of
particular importance for the GVDM approach (for a review of GVDM models, see \cite{donnachie-shaw}). Furthermore, the GVDM
model based on hadronic fluctuations of the photon is natural
to combine with our model \cite{EI} for hadronic fluctuations of the
target proton, which has been used to derive the non-perturbative
$x$-shape of the proton's parton density functions. Combining
parton density functions including DGLAP evolution \cite{DGLAP} with
GVDM gives a good description of data over the full $Q^2$
region. This extends earlier work \cite{badelek,sch-sp} on applying
GVDM and is complementary to theoretical developments where GVDM is connected with a QCD dipole approach \cite{Nikolaev:1990ja,Frankfurt:1997zk,Cvetic:2001ie,Kuroda:2003np}.

\section{Vector meson dominance model for $ep$ at low $Q^2$}
The occurence of quantum fluctuations implies that a photon may also appear as a vector meson such that the quantum state should be expressed as
\begin{equation}
\label{eq:photon-fluctuation}
|\gamma\rangle = C_0|\gamma_0\rangle + \sum_V \frac{e}{f_V}|V\rangle + \int_{m_0}dm (\cdots)
\end{equation}
The first vector meson dominance model included only the sum
over the vector meson states $V = \rho^0, \omega, \phi \ldots$,
whereas the generalised model \cite{GVDM} also includes the integral
over a continuous mass spectrum (not written out explicitly in
eq.~(\ref{eq:photon-fluctuation})).

This hadronic fluctuation of the photon then interacts with the target proton with a normal hadronic cross section dominated by soft processes without any hard scale involved. Total cross sections for different beam hadrons at different energies are well measured and given by standard parametrisations to be discussed below. The overall cross section is then a convolution of the photon-to-meson fluctuation probability with the meson propagator and the meson-proton cross section. 

In $ep$ scattering\footnote{The DIS variables are
defined through $Q^2=-q^2=-(p_e-p_e^\prime)^2$, $x=Q^2/2P\cdot q$,
$y=P\cdot q/P\cdot p_e$ in terms of the four-momenta $P,p_e,
p_e^\prime, q$ of the incoming proton, incoming and scattered electron
and the exchanged photon, respectively.} data is given in terms of the proton structure function $F_2$ extracted from the differential cross section $d\sigma/dxdQ^2$ for electromagnetic interactions (one-photon exchange), since the weak interactions are completely negligible for $Q^2\ll m_{Z,W}^2$. The
structure function $F_2$ can be expressed as \cite{hand,VDM}
\begin{equation}
\label{F2-sigmaTL}
F_2(x, Q^2) = \frac{Q^2 (1-x)}{4\pi^2\alpha \left( 1+4x^2m_p^2/Q^2 \right)}
	[\sigma_T(x, Q^2)+\sigma_L(x, Q^2)]
\end{equation}
in terms of the total cross sections $\sigma_T$ and $\sigma_L$ for transverse and longitudinal virtual photons. 

These cross sections are obtained by squaring the amplitude involving
expression (\ref{eq:photon-fluctuation}) whose continuous part results in a double mass integral 
\linebreak 
$\int_{m_0^2}dm^2d{m'}^2
\frac{\tilde\rho_{T,L}(W^2,m^2,{m'}^2)m^2{m'}^2}{(m^2+Q^2)({m'}^2+Q^2)}$ 
\cite{GVDM}. Off-diagonal contributions having $m\ne m^\prime$ 
\cite{Fraas:gh} are normally neglected in phenomenological studies on 
nucleons, although they cannot be 
neglected for nuclei \cite{Nikolaev:1990ja,Shaw}. Since we here only
consider nucleons, we take this integral to be diagonal, {\it i.e.}\
$\tilde\rho_{T,L}(W^2,m^2,{m'}^2) =
\rho_{T,L}(W^2,m^2)\delta(m^2-{m'}^2)$. The spectral weight function
$\rho_T(W^2,m^2)$ is phenomenologically chosen to fit data, {\it e.g.}\
$\rho_T=m_0^2/m^4$ to obtain scaling at larger $Q^2$, while $\rho_L=\xi_C\frac{Q^2}{m^2}\rho_T$. In this GVDM approach, the resulting cross-sections are \cite{GVDM}
\begin{eqnarray}\label{sigmaTL-GVDM}
\sigma_T^\textrm{\tiny GVDM} &=& \sum_{V} \frac{4\pi\alpha}{f_V^2}
	\left(\frac{m_V^2}{Q^2+m_V^2}\right)^2 \sigma_{V p}
	\;\; + \;\; \frac{m_0^2}{Q^2+m_0^2 }\sigma_{C p}
	\\
\sigma_L^\textrm{\tiny GVDM} &=& \sum_{V} \frac{4\pi\alpha}{f_V^2}
	\frac{Q^2}{m_V^2}
	\left(\frac{m_V^2}{Q^2+m_V^2}\right)^2 \xi_V \sigma_{V p}
	\nonumber \\
 & & 	\;\; + \;\; \left(\frac{m_0^2}{Q^2}
	\ln\left(1 + \frac{Q^2}{m_0^2}\right) - 
	\frac{m_0^2}{Q^2+m_0^2}\right)\xi_C \sigma_{C p}
\end{eqnarray}
In the sums over the discrete vector meson states one recognises the well-known factors $4\pi\alpha/f_V^2$ (involving the vector meson decay constant $f_V$) which give the probabilities of the fluctuations $\gamma \to V$ for real photons, followed by the squared propagator of the meson with mass $m_V$ and the meson-proton total cross section $\sigma_{V p}$. The terms proportional to $\sigma_{C p}=r_C\, \sigma_{\gamma p}$ (defined exactly below) originates from the integral over the continuous vector meson mass spectrum with a lower limit given by the parameter $m_0$. The parameters $\xi_V = \sigma^L_{Vp}/\sigma^T_{V p}$ and $\xi_C = \sigma^L_{C p}/\sigma^T_{C p}$ accounts for the possibility of different cross sections for transverse and longitudinal polarisation states. It is assumed that they are independent of $x$ and $Q^2$ and expected that they are less than unity. 

The total cross-sections $\sigma_{V p}$ and $\sigma_{\gamma p}$ can be directly taken as the well known and generally used parametrisation \cite{DL} 
\begin{equation}\label{sigma-total}
\sigma(ip\rightarrow X) = A_i s^\epsilon + B_i s^{-\eta}
\end{equation}
for the total cross section of a particle $i$ on a proton. The first term is for pomeron exchange and the second one for reggeon exchange. The energy dependence is given by the parameters $\epsilon\approx 0.08$ and $\eta\approx 0.45$ which are universal and obtained from fits to a wealth of data on total cross sections, whereas the normalisation parameters $A_i,B_i$ are different for different particles. At high energies the reggeon term can be neglected in comparison to the dominating pomeron term.

This parametrisation applies not only to the vector mesons ($i=V$) but also to photons ($i=\gamma$) which are on-shell or nearly so. Thus we have $\sigma_{V p} = A_V s_\gamma^\epsilon + B_V s_\gamma^{-\eta}$ and $\sigma_{\gamma p} = A_\gamma s_\gamma^\epsilon + B_\gamma s_\gamma^{-\eta}$. The fractions of the $\gamma p$ cross section accounted for by the discrete vector mesons $V$ are then $r_V=\frac{4\pi\alpha}{f_V^2}\cdot \frac{A_V}{A_\gamma}$, and we can specify $r_C = 1- \sum_V r_V$ as the fraction from the continuous mass spectrum.

Inserting these GVDM expressions for $\sigma_{T,L}$ in eq.~(\ref{F2-sigmaTL}) one obtains
\begin{eqnarray}\label{F2-GVDM}
F_2(x,Q^2) & = & \frac{(1-x)Q^2}{4\pi^2\alpha} 
      \left\{  \sum_{V=\rho, \omega, \phi} r_V \left(\frac{m_V^2}{Q^2 + m_V^2}\right)^2
	\left(1 + \xi_V\frac{Q^2}{m_V^2}\right) \right.
\nonumber \\
 & & \left. +\; r_C\left[ (1-\xi_C)\frac{m_0^2}{Q^2 + m_0^2} + 
        \xi_C \frac{m_0^2}{Q^2}\ln{(1 + \frac{Q^2}{m_0^2})}
	\right] \right\}
	A_\gamma \frac{Q^{2\epsilon}}{x^\epsilon}
\end{eqnarray}
where the following approximations, which are justified for the region of $x$ and $Q^2$ of HERA data, have been made: In the prefactor the term  $4x^2m_p^2/Q^2\ll 1$ and is hence neglected. The last factor originating from $\sigma_{Vp}$ and $\sigma_{Cp}$ only includes the pomeron term, since the reggeon term is negligible, and the energy variable is   
$s_{\gamma p} = Q^2\: \frac{1-x}{x} + m_p^2 \approx Q^2/x$ at small-$x$.

The parameters involved in eq.~(\ref{F2-GVDM}) are all essentially known from GVDM phenomenology. The values $r_{V=\rho,\omega,\phi,C} = 0.67, 0.062, 0.059, 0.21$ are quite well determined  \cite{VDM}. Although $m_0\approx 1$ GeV is expected \cite{sch-sp}, it is not well known and is here taken as a free parameter. The parameters $\xi_V$ are assumed to be the same for $V=\rho,\omega,\phi$ and expected to be $\xi_V\approx 0.25$ based on the early study in \cite{GVDM} and supported by \cite{Kuroda:2003np} including recent HERA data. A similar magnitude is expected for $\xi_C$. Lacking established numbers and wanting to have as few parameters as possible, we use the common parameter $\xi=\xi_V=\xi_C$ as a free parameter to be fitted. For the pomeron intercept parameter the value $\epsilon=0.09$ has been obtained in recent fits \cite{cudell}, but we take it as a free parameter in order to check the expected consistency with this universal value. Also the overall normalisation constant $A_\gamma$ of the photon-proton cross section is taken as a free parameter. Thus, we have the four parameters $\xi,m_0,\epsilon,A_\gamma$ to be fitted to data.

\section{Comparison to $F_2$ data}
\begin{figure}[thbp]
\begin{center}
\includegraphics*[width=11cm]{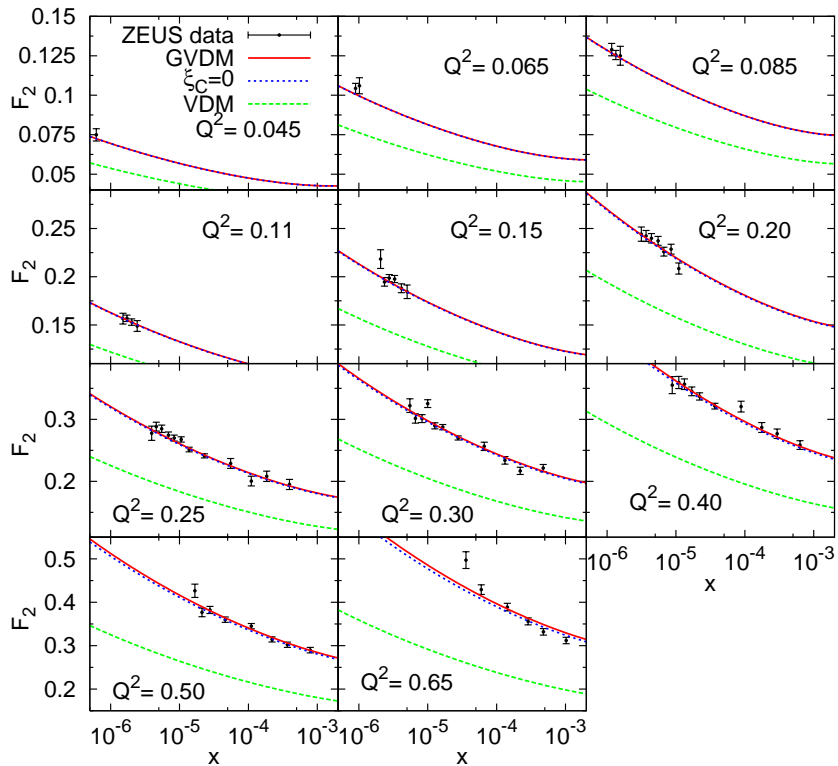}
\caption{$F_2$ at low $Q^2$: HERA $ep$ data from ZEUS \cite{ZEUS} compared to GVDM as in eq.~(\ref{F2-GVDM}) (full curves). Model results are also given when the longitudinal contribution of the continuum is excluded ($\xi_C=0$) and when excluding the continuous contribution altogether (setting $r_C=0$) giving VDM.}
\label{fig:lowQ2}
\end{center}
\end{figure}

The GVDM expression for $F_2$ in eq.~(\ref{F2-GVDM}) gives a very good description of the HERA data on $F_2$ at low $Q^2$, as shown in Fig.~\ref{fig:lowQ2}. The fit gives $\chi^2/\rm{d.o.f.} = 87/(70-4) = 1.3$ with parameter values as expected: $\epsilon = 0.091$, $\xi = 0.34$, $m_0=1.5$~GeV just above the discrete vector meson masses and $A_\gamma = 71\, \mu\rm{b}$ in accordance with the measured photon-proton cross section (cf.\ \cite{sch-sj}). This demonstrates that for $Q^2$ clearly below 1 GeV$^2$ the HERA $ep$ cross section can be fully accounted for by GVDM using parameter values as determined from old investigations related to fixed target data. 

\begin{figure}[thbp]
\begin{center}
\includegraphics*[width=8cm]{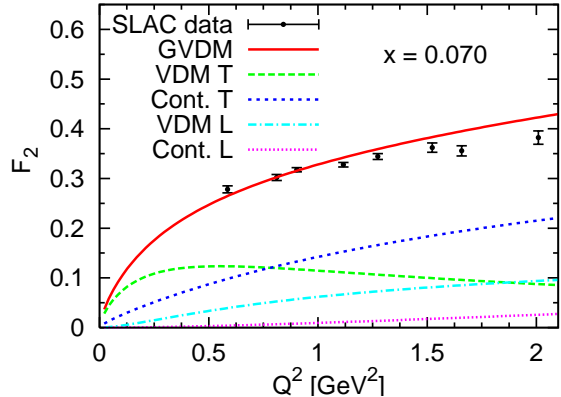}
\caption{The $Q^2$ dependence of $F_2$ from GVDM (full curve) with its contributions from transverse (T) and longitudinal (L) parts of the discrete vector meson spectrum (VDM) and the continuous (Cont.) mass spectrum. Data from SLAC \cite{slac} are included for comparison.}
\label{fig:SLAC}
\end{center}
\end{figure}

For completeness, both the transverse and longitudinal contributions to the integral over the continuous mass spectrum are here included, although the latter is numerically small as demonstrated in Fig.~\ref{fig:lowQ2}. VDM, which lacks the continuum part, falls below the data and decreases too fast with $Q^2$. This $Q^2$ behaviour becomes even worse if the longitudinal contribution is neglected ({\it i.e.}\ $\xi_V=0$), as is done in some simplified treatments of VDM. The $Q^2$ dependence of these different contributions is shown in Fig.~\ref{fig:SLAC}.

We have also compared with data on $F_2$ from SLAC \cite{slac} and NMC \cite{nmc}. Due to the lower energies of these fixed target experiments, one must here include also the reggeon term in the Donnachie-Landshoff parameterisation of the total cross section and we use $\eta=0.45, B_\gamma=90\, \mu\rm{b}$ ({\it cf.}\ \cite{sch-sj,DL}). Keeping the values of the other parameters fixed, we obtain good agreement as long as $x$ and $Q^2$ are not too large ({\it cf.}\ \cite{Donnachie:2001xx}). 

At larger $Q^2$, this original GVDM does not have the correct
behaviour since $F_2$ in eq.~(\ref{F2-GVDM}) increases with $Q^2$ for
all $x$. This can be cured
phenomenologically by introducing for the spectral weight function mentioned above a suitable form $\rho_T = N \ln{(W^2/am^2)}/m^4$ \cite{sch-sp}. With
suitable values of the free parameters $m_0,N,a$ it is then possible
to reproduce HERA $F_2$ data also at larger $Q^2$. A theoretically
more advanced alternative is to instead include off-diagonal
contributions \cite{Nikolaev:1990ja,Cvetic:2001ie}. This connects naturally to the dipole formalism of DIS and include effects of perturbative QCD
evolution. This off-diagonal GVDM framework should then apply  in
the full $Q^2$ region, as long as $x$ is sufficiently small, and HERA data can here be reproduced \cite{Cvetic:2001ie}. 

At high $Q^2$ the conventional description is in terms of parton density functions, which also includes the large-$x$ valence region. As argued above, this approach does not apply at very small $Q^2$ and one must therefore complement it with GVDM to account for this region. To cover the full $x$ and $Q^2$ region one should combine these two descriptions, but due to the confinement problem, there is no proper theoretical way to do the transition from GVDM formulated in a hadron basis to the parton model in a parton basis. Although GVDM can be extended to large $Q^2$, this would imply double counting if combined with the conventional parton description. To use the latter one must, therefore, phase out GVDM.

Thinking in terms of the resolution scale discussed above, it is quite
natural that the original hadron-based GVDM only applies at low $Q^2$
and there should be a transition to the DIS formalism of resolved
partons at high $Q^2$. In particular, the total cross sections
$\sigma_{Vp},\sigma_{Cp}$ used in GVDM applies to soft hadronic
processes for (nearly) on-shell particles. It is therefore very
reasonable to phase out GVDM at larger $Q^2$ by applying a form factor
suppression. A factor like $m_V^2/(m_V^2 + Q^2)$ \cite{f-sj} would,
however, ruin the very good description at low $Q^2$ seen in
Fig.~\ref{fig:lowQ2}. Instead, a sharper transition to DIS in the
region $Q^2 = 0.6 - 1.5 \rm{\,GeV}^2$ is required. This is in
accordance with the rather abrupt change of the slope parameter
$\lambda$ in $F_2(x)\sim x^{-\lambda}$ observed in HERA data at
$Q^2\approx 1\, \rm{GeV}^2$ \cite{lambda} and may be seen more
generally as a rather sharp transition from soft, non-perturbative to
hard, perturbative QCD dynamics.

\begin{figure}[hbtp]
\begin{center}
\includegraphics*[width=10cm]{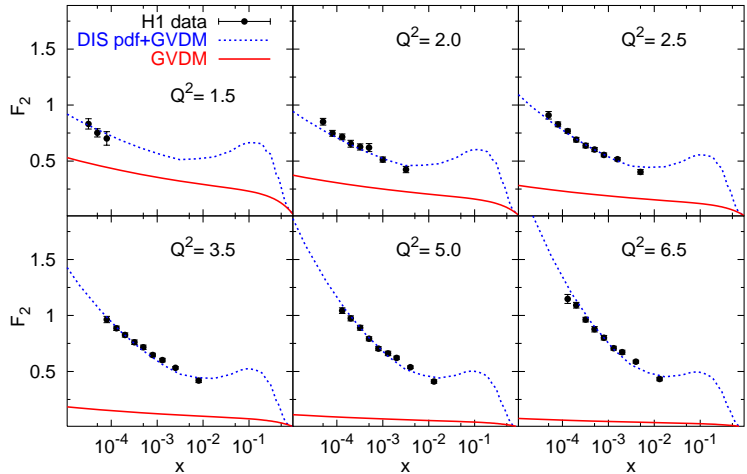}
\caption{$F_2$ at intermediate $Q^2$: contribution of GVDM with a form factor $(1.24/Q^2)^{1.63}$ (full curve) and the complete model (dashed curve), including also DIS parton density functions from our model, compared to H1 data \cite{H1}.}
\label{fig:intermediateQ2}
\end{center}
\end{figure}

We therefore introduce the phenomenological form factor
$(Q^2_C/Q^2)^a$ for $Q^2>Q^2_C$ to phase out GVDM above a critical
$Q^2_C$.  As shown in Fig.~\ref{fig:intermediateQ2}, a good
description of HERA $F_2$ data at intermediate $Q^2$ can then be
obtained by combining GVDM and parton density functions that fit HERA
$F_2$ data at larger $Q^2$. This requires $Q^2_C\approx 1\,
\rm{GeV}^2$ as expected from the discussed transition, and $a\approx
2$ giving $\sim Q^{-4}$ as a reasonable form factor damping. 
The exact values of the parameters are fitted and
depend on the details of the DIS parton densities. With such a form
factor suppression, the GVDM contribution is negligible for
$Q^2\gtrsim 4\, \rm{GeV}^2$ (see fig.~\ref{fig:intermediateQ2}), where
DIS parton density parametrisations are usually considered
trustworthy. Any parametrisation of parton densities which is good
enough to reproduce the measured $F_2$ in the DIS region can be used,
provided the GVDM component is taken into account when low-$Q^2$
data are included in the fits. 

For Fig.~\ref{fig:intermediateQ2} we have, however, used a physically motivated model \cite{EI} where the parton momentum distributions are obtained from gaussian fluctuations having widths related to the uncertainty relation and the proton size. Valence distribtuions arise from the `bare` proton, whereas sea distributions originate from mesons in hadronic fluctuations of the proton $|p\rangle  =  \alpha_0|p_0\rangle + \alpha_{p\pi}|p_0\pi^0\rangle + \alpha_{n\pi}|n \pi^+\rangle + \ldots + \alpha_{\Lambda K}|\Lambda K^+\rangle + \ldots$
This gives the $x$-shape of the parton densities at $Q^2_0\approx 1\, \rm{GeV}^2$ and the DGLAP equations are then used to evolve to larger $Q^2$,
resulting in a good fit to HERA $F_2$ data using only six parameters with physically motivated values \cite{EI}. Furthermore, this model gives \cite{AI} $u_v(x)\ne d_v(x)$ and $\bar{u}(x)\ne \bar{d}(x)$ in qualitative agreement with data, as well as $s(x)\ne \bar{s}(x)$ of interest for the NuTeV anomaly \cite{nutev}. 

It is interesting that combining these models involving quantum fluctuations of both the photon and the target proton results in a good description of the $ep$ cross section, or equivalently $F_2$, at both low and high $Q^2$. 

\section{Conclusions} 
The conventional parton model formulation of deep inelastic scattering
is not applicable at very low $Q^2$, where no hard scale is available
to resolve the partons. Instead, HERA $F_2$ data are here well
reproduced by the original generalised vector meson dominance model,
including contributions from a continuous mass spectrum and
longitudinal polarisation states, and using parameter values in
agreement with old analyses at fixed target energies. 
At large $Q^2$, GVDM with off-diagonal contributions can be used as long as $x$ is small. To cover the full $x$-region, including the valence part, the proton structure must be introduced via parton density functions in the conventional DIS formalism. 

We have shown that one can combine the GVDM and parton
density descriptions in a two-component phenomenological model. GVDM
then accounts fully for the cross section below $Q^2\lesssim 1$~GeV, but although it contributes also at large $Q^2$ it must here be phased out in order to avoid double counting with the standard parton density formulation. We have found that a form factor damping of GVDM gives a smooth transition into the deep inelastic region described by parton distribution functions. Here,
any good parametrisation of parton densities can be used, provided the
GVDM component is taken into account at low $Q^2$ as shown above when fitting the parameters. In this way one obtains a
good overall result at both low and high $Q^2$. In particlar, there is
no need for a negative gluon density in the region of low $x$ and low $Q^2$. 
The reason is that the cross section is here dominated by the GVDM
contribution, which is based on fundamental quantum fluctuations that should 
not be neglected.

{\bf Acknowledgments:} We are grateful to Dieter Schildknecht for interesting and helpful discussions and to Johan Rathsman for a critical reading of the manuscript.


\begin{thebibliography}{22}

\bibitem{negative-gluon} 
A.D.\ Martin {\it et al.}, Eur.\ Phys.\ J.\ {\bf C23} (2002) 73;\\
J.\ Pumplin {\it et al.}, JHEP {\bf 0207} (2002) 012.

\bibitem{mrst} 
A.~D.~Martin {\it et al.}, 
Eur.\ Phys.\ J.\ C {\bf 28} (2003) 455; 
arXiv:hep-ph/0308087.

\bibitem{VDM} T.H.\ Bauer {\it et al.}, Rev.\ Mod.\ Phys.\ {\bf 50} (1978) 261.

\bibitem{GVDM} J.J.\ Sakurai and D.\ Schildknecht, Phys.\ Lett.\ {\bf B40} (1972) 121.

\bibitem{donnachie-shaw} A.\ Donnachie and G.\ Shaw in {\em Electromagnetic Interactions of Hadrons}, vol.~2, 
editors A.\ Donnachie and G.\ Shaw, Plenum Press, New York, 1978.

\bibitem{EI} A.\ Edin and G.\ Ingelman, Phys.\ Lett.\ {\bf B432} (1998) 402;
 Nucl.\ Phys.\ Proc.\ Suppl.\ B {\bf 79} (1999) 189.

\bibitem{DGLAP}
V.~N.~Gribov and L.~N.~Lipatov, Sov.\ J.\ Nucl.\ Phys.\  {\bf 15}, 438 (1972);\\ 
G.~Altarelli and G.~Parisi, Nucl.\ Phys.\ B {\bf 126}, 298 (1977);\\
Yu.~L.~Dokshitzer, Sov.\ Phys.\ JETP {\bf 46}, 641 (1977).

\bibitem{badelek}
B.~Badelek and J.~Kwiecinski, Rev.\ Mod.\ Phys.\  {\bf 68} (1996) 445
and references therein. 

\bibitem{sch-sp}
D.~Schildknecht and H.~Spiesberger,
arXiv:hep-ph/9707447.

\bibitem{Nikolaev:1990ja}
N.~N.~Nikolaev and B.~G.~Zakharov,
Z.\ Phys.\ C {\bf 49} (1991) 607.

\bibitem{Frankfurt:1997zk}
L.~Frankfurt, V.~Guzey and M.~Strikman,
Phys.\ Rev.\ D {\bf 58} (1998) 094039.

\bibitem{Cvetic:2001ie}
G.~Cvetic, D.~Schildknecht, B.~Surrow and M.~Tentyukov,
Eur.\ Phys.\ J.\ C {\bf 20} (2001) 77.

\bibitem{Kuroda:2003np}
M.~Kuroda and D.~Schildknecht,
arXiv:hep-ph/0309153.

\bibitem{hand} L.N.\ Hand, Phys.\ Rev.\ {\bf 129} (1963) 1834.

\bibitem{Fraas:gh}
H.~Fraas, B.~J.~Read and D.~Schildknecht,
Nucl.\ Phys.\ B {\bf 86} (1975) 346.

\bibitem{Shaw}
G.~Shaw,
Phys.\ Lett.\ B {\bf 228} (1989) 125;
Phys.\ Rev.\ D {\bf 47} (1993) 3676.


\bibitem{DL} A.~Donnachie and P.V.~Landshoff, Phys.\ Lett.\ B {\bf 296} (1992) 227.

\bibitem{cudell} 
J.~R.~Cudell {\it et al.}, Phys.\ Rev.\ D {\bf 61} (2000) 034019
[Erratum-ibid.\ D {\bf 63} (2001) 059901];\\
J.~R.~Cudell {\it et al.}, Phys.\ Lett.\ B {\bf 395} (1997) 311.

\bibitem{sch-sj} G.A.\ Schuler and T.\ Sjöstrand, Nucl.\ Phys.\ B {\bf 407} (1993) 539.

\bibitem{ZEUS} 
J.~Breitweg {\it et al.}\ [ZEUS Collaboration],
Phys.\ Lett.\ B {\bf 487} (2000) 53.

\bibitem{slac} L.W.\ Whitlow {\it et al.}, Phys.\ Lett.\ {\bf B282} (1992) 475 
(data on $F_2^p$ obtained from the collaboration).

\bibitem{nmc} M.\ Arneodo {\it et al.}, Nucl.\ Phys.\ {\bf B483} (1997) 3.

\bibitem{Donnachie:2001xx}
A.~Donnachie and P.~V.~Landshoff,
Phys.\ Lett.\ B {\bf 518} (2001) 63.

\bibitem{f-sj} C.\ Friberg and T.\ Sjöstrand, JHEP {\bf 0009} (2000) 010.

\bibitem{lambda} J.\ Breitweg {\it et al.}, Eur.\ Phys.\ J.\ {\bf C7} (1999) 609.

\bibitem{H1} C.\ Adloff {\it et al.}, Eur.\ Phys.\ J.\ {\bf C21} (2001) 33.

\bibitem{AI} J.\ Alwall and G.\ Ingelman, in preparation.

\bibitem{nutev} G.~P.~Zeller {\it et al.}\ [NuTeV Collaboration], Phys.\ Rev.\ Lett.\ {\bf 88} (2002) 091802.

\end{thebibliography}
\end{document}